\documentclass{article}
\usepackage{graphicx}
\usepackage{amsmath}
\usepackage{amsfonts}
\usepackage{amssymb}
\begin{document}

\title{Relevance of the Inherent Structures and Related Fundamental Assumptions in
the Energy Landscape}
\author{P.D. Gujrati and F. Semerianov\\The Department of Physics and The Department of Polymer\ Science, \\The University of Akron, Akron, Ohio 44325.}
\date{\today}
\maketitle
\begin{abstract}
We carefully investigate the two fundamental assumptions in the
Stillinger-Weber analysis of the inherent structures (IS's) in the energy
landscape and come to conclude that they cannot be validated. This explains
some of the conflicting results between their conclusions and some recent
rigorous and exact results. Our analysis shows that basin free energies, and
not IS's, are useful for understanding glasses.
\end{abstract}

It is well known that most supercooled liquids (SCL) become viscous when their
configurational entropy $S(T)$ \cite{Note1} becomes negligible as they are
cooled, provided the corresponding crystal (CR) is not allowed to
nucleate.\ The flow practically ceases over a period constrained by
experimental limits, the viscosity becomes very large, and the viscous fluid
eventually becomes an amorphous solid or glass. Our current understanding of
glassy behavior is still far from complete, even after many decades of
continuous investigation. In order to better understand the flow properties of
viscous fluids, Goldstein proposed the potential energy landscape picture
using classical statistical mechanical\emph{\ canonical ensemble
}\cite{Goldstein}, in which the energy barriers control the flow at low
temperatures. The discussion was mostly qualitative, but provided an
interesting and sufficiently tractable scheme and included some quantitative
predictions in the SCL and the glassy states. Stillinger and Weber (SW) later
revived this picture, carried out an analysis in terms of basins, and
concluded that their minima, called the inherent structures (IS), play a
pivotal role in the thermodynamics of viscous fluids at low temperatures
\cite{Stillinger,Stillinger1}. The IS-entropy $S_{\text{IS}}(T)$ [not to be
confused with $S(T)]$ of the IS's vanishes at some $T=$ $T_{\text{SW}}$
(called $T_{\text{K}}$ by SW, but we will reserve $T_{\text{K}}$ where
$S(T)=0$ $\cite{Note1}$) so that the system gets trapped into a single basin
at $T_{\text{SW}}$ so that $S(T)$ becomes the basin configurational entropy
$S_{\text{b}}(T)\geq0$ below $T_{\text{SW}}\geq T_{\text{K}}\cite{Note1}.$ The
SW analysis has given rise to a considerable amount of literature\ in recent
years; for a partial list, see \cite{ISliterature}.

One of the most tantalizing consequences of the SW analysis is the theoretical
conclusion $T_{\text{SW}}=0$ drawn by Stillinger\cite{Stillinger1}.
Consequently, it is not possible to have a configurational entropy crisis
($S(T)<0$) below a positive temperature $T_{\text{K}}$ \cite{Note1} for SCL.
For CR, the claim implies that just above $T_{\text{SW}}=0,$ CR probes many
basins ($S_{\text{IS}}>0$)$,$ not all of which have to be close in the
configuration space, and its heat capacity is \emph{not} due to pure
vibrations within a single basin. This is hard to understand in view of the
tremendous success of the Debye model. Stillinger's argument also does not
permit any SCL spinodal at a positive temperature. However, the latter has
been observed in exact calculations for finite-length polymers \cite{GujRC}
and in a binary mixture $\cite{Gujrati0}$. The theoretical conclusion
$T_{\text{SW}}=0$ \cite{Stillinger1} is contradicted by numerous numerical
evidence \cite{ISliterature} of a positive $T_{\text{SW}}$ within the IS
picture. A recent rigorous statistical mechanical proof by Gujrati
$\cite{Gujrati}$ also contradicts this claim. He has shown that under the
assumption that the ideal glass has a higher energy ($E=E_{\text{K}}$) than
the corresponding crystal ($E=E_{0}<E_{\text{K}}$)\ at absolute zero, $S(T)$
of the stationary SCL (obtained under infinitely slow cooling of the
disordered equilibrium liquid EL) must necessarily vanish at $T_{\text{K}}>0$.
At $T_{\text{K}}$, SCL has its configurational energy $E_{\text{K}}.$ Two
independent proofs are given $\cite{Gujrati},$ and the conclusions are also
substantiated by two exact model calculations, one of which is not mean-field.
Other exact calculations for the abstract random energy model \cite{Derrida},
and for long polymers \cite{GujRC,Guj} also support the conclusions by
Gujrati. In addition, a recent exact solution by Semerianov and
Gujrati\cite{Fedor} of a dimer model, a prototype model of molecular liquids,
also exhibits configurational entropy crisis in its stationary metastable
state below a positive temperature. Corsi \cite{Corsi} \ has also observed
positive Kauzmann temperatures in exact calculations for small
particles\ occupying four and fives lattice sites each. The important point to
note is that none of these exact calculations utilizes the energy landscape.
However, if the latter has any validity, its consequences must be in
accordance with the exact calculations and the rigorous analysis, which is
certainly not the case. Thus, we need to reexamine the SW analysis to clarify
the conflict. We also provide an alternative analysis of the landscape, which
is consistent with the rigorous analysis.

The canonical ensemble free energy $F_{\text{dis}}(T)$ of the disordered EL is
\emph{continued} analytically \cite{Gujrati} below the melting temperature
$T_{\text{M}}$ to give the SCL\ free energy. As a mathematical continuation,
the resulting SCL does not have to satisfy the reality condition $S(T)\geq0. $
The continuation, in principle, can stop in a spinodal singularity at a
positive temperature \cite{GujRC,Gujrati0}. Here, we are only interested in
the case when the SCL free energy can be mathematically continued all the way
down to $T=0$ without encountering any singularity$.$ It is easy to show
\cite{Gujrati} that, at $T=0$, the SCL and CR free energies are
\emph{identical }$(F=E_{0}),$ provided\ $TS(T)\rightarrow0$. They are again
equal at $T_{\text{M}}$ because of which SCL is forced to exhibit the entropy
crisis ($S(T)<0$) \cite{Gujrati} below $T_{\text{K}}>0$
$\cite{GujRC,Gujrati,Guj},$ where it is discarded and replaced by an ideal
glass phase to \emph{satisfy} the reality condition. The ideal glass has
$E=E_{\text{K}}$ for $T\leq$ $T_{\text{K}}$. At $T=0$, the configuration
corresponding to $E_{\text{K}}$ must represent a potential energy minimum,
provided we neglect surface effects. At a positive temperature $T_{\text{S}%
}>T_{\text{K}}$, SCL gets confined into the basin whose IS is at $E_{\text{K}%
}.$ Thermal fluctuations will still allow SCL to visit other basins, which are
rare if $T$ is small. We will not investigate fluctuations and restrict
ourselves only to the average behavior here.

The following two observations \cite{Gujrati} are going to be relevant below.

G1. The zero of the temperature scale is set by the global potential energy
minimum $E_{0}.$

G2. The ideal glass configuration at $E_{\text{K}}$ is a local minimum of the
potential energy (neglecting surface effects), which SCL\ approaches at
$T_{\text{K}}>0$. Since SCL \emph{does not} physically exist below
$T_{\text{K}}$ (due to negative entropy), the ideal glass is put in by hand
for $T<T_{\text{K}}$ to avoid the entropy crisis. It does not emerge directly
in the statistical mechanical description.

In continuum classical statistical mechanics, the dimensionless total
partition function (PF) $Z_{\text{T}}$ of a system of $N$ identical particles
$i=1,2,...,N$ in a given volume $V$,\ with the Hamiltonian $E_{\text{T}%
}(\{\mathbf{r}_{i},\mathbf{p}_{i}\})\equiv K(\{\mathbf{p}_{i}%
\})+E(\{\mathbf{r}_{i}\}),$ $K,E$ being the kinetic and the configurational
(i.e., the potential) energy, is a product of two \emph{independent} integrals

$\qquad\qquad\qquad\qquad\qquad\qquad\qquad\qquad\qquad\qquad\qquad
\qquad\qquad\qquad\qquad$%
\begin{equation}
Z_{\text{T}}\equiv\frac{1}{(2\pi\hslash)^{3N}}\int e^{-\beta K}d^{N}%
\{\mathbf{p\}\int^{^{\prime}}}e^{-\beta E}d^{N}\mathbf{\{r\}.}\label{PF}%
\end{equation}
Here $d^{N}\{\mathbf{p\}}$,$d^{N}\{\mathbf{r\}}$ represent integrations with
respect to momenta and positions $\mathbf{p}_{i},\mathbf{r}_{i}$ of the
particles, and $\beta$ the inverse temperature $1/T$ in the units of the
Boltzmann constant $k_{\text{B}}.$ The second integral $\mathbf{(}$the prime
implying integration over \emph{distinct} configurations of the particles) is
called the configurational PF, to be denoted by $Z$.

The momentum integration in $Z_{\text{T}}$ can be expressed in terms of
$W_{\text{KE}}(P)dP$ $\equiv C_{3N}P^{3N-1}dP/h^{3N}$ related to the
3$N$-dimensional momentum space volume within the spherical shells of radii
$P$, and $P+dP$ [$C_{d}\equiv$ $d\pi^{d/2}/\Gamma(d/2+1)].$ The translational
entropy due to the translational degrees of freedom is given by $S_{\text{KE}%
}(T)=\ln W_{\text{KE}}(\overline{P}),$ $\overline{K}\equiv\overline{P}%
^{2}/2m=3NT/2.$ In the thermodynamic limit $N\rightarrow\infty$, we find that
$S_{\text{KE}}(T)=(3N/2)[1+\ln T+\ln(2\pi m/h^{2})],$ and has the \emph{same
value} at a given temperature for \emph{all} classical systems, regardless of
their configurational energy. Thus, in general, the entropy due to the
configurational degrees of freedom can be always obtained by subtracting
$S_{\text{KE}}(T)$ from $S_{\text{T}}(T)\cite{GujRC,Fedor}:S(T)\equiv
S_{\text{T}}(T)-S_{\text{KE}}(T),$ where $S_{\text{T}}(T)$ is the total
entropy in the canonical ensemble; see (\ref{PF}). For an ideal gas, $Z$
$=V^{N}/N!,$ so that $S=N\ln(Ve/N)$, which no longer depends on $T. $ For
$T<eh^{2}/2\pi m$, $S_{\text{KE}}(T)<0$, or for $V/N<1/e$, $S(T)<0.$

The problem of negative entropy is well known in classical statistical
mechanics, and requires quantum statistical mechanics for its resolution.
Unfortunately, it is not possible to solve a quantum statistical mechanical
model exactly at present.\ Thus, care must be exercised when drawing
conclusions based on the sign of the entropy. Since $T_{\text{K}}$ is signaled
by a negative entropy, it is \emph{crucial} to have a formalism in which the
entropy is never negative for realizable states in Nature
\cite{GujRC,Gujrati,Guj}. The simplest way to achieve this is to discretize
either the phase space by using cells of size $h^{3N}$ or the real space by
using a lattice$.$ Thus, in the following, we take it for granted that such a
discretization has been carried out$.$ We closely follow \cite{Stillinger1}
who factors out the kinetic energy part and only uses $Z$ to introduce the
potential landscape picture\cite{Goldstein,Stillinger}, and consider only the
configurational entropy $S(T)$ \cite{Note1}, the canonical PF $Z(T)$, and the
configurational free energy is $F(T)\equiv-T\ln Z(T)$ in the following. We
rewrite $Z(T)$ as follows. Let $W(E)\geq1$ (so that $S(E)\equiv\ln W(E)\geq0$
\cite{GujRC,Gujrati,Guj} for physically realizable states) denote the
\emph{number} of the configurations of potential energy $E.$ Then $Z(T)$ can
be rewritten as a sum over $E:$ $Z(T)\equiv\sum_{E}e^{-\beta E}W(E)$.
Replacing $W(E)$ in $Z(T)$ by the number of disordered configurations
$W_{\text{dis}}(E),$ we obtain the PF\ $Z_{\text{dis}}(T)$ so that
$F_{\text{dis}}(T)\equiv-T\ln Z_{\text{dis}}(T)$ is the free energy of the
disordered phase EL and its extension SCL
\cite{Stillinger,Stillinger1,Gujrati}. In the following, we will use $Z(T)$ to
represent both PF's, which should cause no confusion since the context will be clear.

\textbf{SW Picture.} The potential energy landscape is a union of disjoint
basins. A basin\ is indexed by $j$, and characterized by its minimum and
maximum energies $E_{j}$, and $E_{j\text{,max}}$ so that it does not exist
outside this energy range $\Delta_{j}E\equiv$ $(E_{j},E_{j\text{,max}})$. Let
$W_{j}(E)$ ($E\in\Delta_{j}E$) represent the number of \emph{distinct} states
of energy $E$ in the $j$-th basin. We introduce the \emph{shifted} PF%

\begin{equation}
z_{j}(T)\equiv\underset{E}{\sum}W_{j}(E)e^{-\beta(E-E_{j})}\;\widehat{\delta
}_{E,\Delta_{j}E}\label{ShftPF}%
\end{equation}
of the $j$-th basin$.$ Here, $\widehat{\delta}_{E,\Delta_{j}E}=1\;$if
$E\in\Delta_{j}E,$ and $0\;$if $E\notin\Delta_{j}E.$ We group basins, indexed
by $j(\lambda)$, into inherent structure classes (ISC) $\mathcal{I}_{\lambda}%
$, indexed by $\lambda$, so that all IS's in a class have the same energy
$E=E_{\lambda}.$ The basin in a class do \emph{not} have to be close in the
configuration space. Let $N_{\text{IS}}(E_{\lambda})$ be the number of basins
in $\mathcal{I}_{\lambda},$ and $S_{\text{IS}}(E_{\lambda})\equiv\ln
N_{\text{IS}}(E_{\lambda})$. Let $Z_{\lambda}\equiv\sum_{j\in j(\lambda)}%
z_{j}(T)$ and $z_{\lambda}\equiv Z_{\lambda}/N_{\text{IS}}(E_{\lambda})$
denote the shifted and the average shifted $\mathcal{I}_{\lambda}$-PF, so
that
\begin{equation}
Z(T)\equiv\sum_{\lambda}e^{-\beta E_{\lambda}}Z_{\lambda}\equiv\sum_{\lambda
}e^{-\beta E_{\lambda}+{}S_{\text{IS}}(E_{\lambda})}z_{\lambda}%
.\label{ISTotPF}%
\end{equation}
Stillinger and Weber \cite{Stillinger,Stillinger1} and various authors in
\cite{ISliterature} assume that $z_{\lambda}$ is an \emph{explicit} function
of $E_{\lambda}$, and $T$ of the form [SW denote quantities that are specific
to the SW-approach]
\begin{equation}
z_{\lambda}^{(\text{SW})}(E_{\lambda},T)\equiv e^{-\beta f^{(\text{SW}%
)}(E_{\lambda},T)}.\label{AvBPF}%
\end{equation}
They also \emph{replace} the first sum over the discrete index $\lambda$ in
$(\ref{ISTotPF})$ by a sum over the almost continuous variable $E_{\lambda}$
so that a general summand can be characterized by $E_{\lambda}$ using
$F^{\text{(SW)}}(E_{\lambda},T)\equiv E_{\lambda}+f^{(\text{SW})}(E_{\lambda
},T)-TS_{\text{IS}}(E_{\lambda}),$ whose minimum with respect to $E_{\lambda}$
at \emph{fixed }$T$ determines $Z(T)$ for a macroscopic system. This minimum
term corresponds to that particular value $E_{\lambda}=\overline{E}_{\lambda}$
at which
\begin{equation}
\left(  \partial S_{\text{IS}}(E_{\lambda})/\partial E_{\lambda}\right)
_{\overline{E}_{\lambda}}=\beta\lbrack1+\left(  \partial f^{(\text{SW}%
)}(E_{\lambda},T)/\partial E_{\lambda}\right)  _{\overline{E}_{\lambda}%
}].\label{ISTMaxCon}%
\end{equation}
The equilibrium free energy, and the IS-entropy are given by $\overline
{F}^{\text{(SW)}}(T)\equiv$ $F^{\text{(SW)}}(\overline{E}_{\lambda},T),$ and
$\overline{S}_{\text{IS}}^{\text{(SW)}}(T)\equiv$ $S_{\text{IS}}(\overline
{E}_{\lambda}),$ respectively. We expect $\overline{S}_{\text{IS}%
}^{\text{(SW)}}(T)$ to vanish at some low $T=T_{\text{SW}}$ and increase
monotonically with $T$ at least at low $T.$ For $T\leq T_{\text{SW}}$, the
system is trapped in a single basin of energy minimum $\overline{E}_{\lambda
}=E_{\text{K}}$ for SCL ($E_{0}$ for CR) so that at or below $T_{\text{SW}}$,
$\overline{E}_{\lambda}$ sticks at $E_{\text{K}}$ for SCL ($E_{0}$ for CR).
Stillinger \cite{Stillinger1} has argued that satisfying (\ref{ISTMaxCon})
\emph{below} $T_{\text{SW}}$ is \emph{inconsistent} with $\overline
{E}_{\lambda}$ sticking at either $E_{\text{K}}$ (or $E_{0}).$ Thus, he
concludes that $T_{\text{SW}}=0,$ a consequence of which is the observation
that there \emph{cannot} be a positive $T_{\text{K}}.$ This conclusion is
based on the following two fundamental assumptions in the SW\ approach:

SW1: The shifted ISC free energy $f^{(\text{SW})}(E_{\lambda},T)-TS_{\text{IS}%
}(E_{\lambda})$ in $F^{\text{(SW)}}(E_{\lambda},T)$ is an explicit function of
$E_{\lambda},$ so that the summation over $\lambda$ can be replaced by that
over $E_{\lambda}.$

SW2: The minimization condition (\ref{ISTMaxCon}) also holds below
$T_{\text{SW}},$ where $\overline{S}_{\text{IS}}^{\text{(SW)}}=0.$

We now demonstrate that neither assumption can be substantiated.

\textbf{Current Analysis}. We observe that $W(E)$ is a \emph{sum} over various
basins: $W(E)\equiv\sum_{j}W_{j}(E)\widehat{\delta}_{E,\Delta_{j}E}.$ We group
all basins having the same minimum at $E=E_{\lambda}$ together into
$\mathcal{I}_{_{\lambda}}.$ It should be noted that $z_{j}(T)$ of all basins
in $\mathcal{I}_{_{\lambda}}$ need not be the identical in value.

1.\qquad\textbf{Evaluating} $\mathit{z}_{j}\mathbf{(\mathit{T})}$\textbf{. }We
now prove that\ $z_{j}$ \emph{cannot} depend explicitly on the basin energy
minimum $E_{j}$, see (\ref{ShftPF}), though it most certainly depends on the
shape of the basin, i.e. on $j$. For example, the curvature of the basin at
its minimum and not its value of $E_{j}$ determines the vibrational
frequencies and the free energy $f_{j}(T)\equiv-T\ln z_{j}$ in the harmonic
approximation$.$ The latter is measured with respect to $E_{j},$ so is
independent of $E_{j}.$ To be sure, let us shift all energies $E\rightarrow
E^{^{\prime}}\equiv E-C$ by some constant $C$ in (\ref{ShftPF})$.$ The number
$W_{j}(E)$ of states, all having the same energy $E,$ remains unchanged under
the shift by $C$. Thus, $W_{j}(E)$ $\rightarrow W_{j}^{\prime}(E^{^{\prime}%
})=W_{j}(E).$ Thus, $z_{j}$ transforms under the shift as $z_{j}%
(T)\rightarrow\sum_{E^{^{\prime}}}W_{j}^{^{\prime}}(E^{^{\prime}}%
)e^{-\beta(E^{^{\prime}}-E_{j}^{^{\prime}})}\widehat{\delta}_{E^{^{\prime}%
},\Delta_{j}E^{^{\prime}}}$ for any arbitrary $C.$ Comparing with
(\ref{ShftPF}), we conclude that $z_{j}$ has not changed. Consequently, it
does not depend on the shift $C,$ including $C=E_{j}.$

Since $W_{j}(E)\geq1$, $z_{j}(T)$ is a sum of \emph{positive} terms. \ Hence,
for a macroscopic system, $z_{j}$ is determined by\ the maximum summand in
(\ref{ShftPF}) corresponding to $E=\overline{E}_{j}\in\Delta_{j}E,$ and the
corresponding heat capacity is non-negative$.$ (Both observations remain valid
even if $W_{j}(E)\geq0,$ a common occurrence in SCL$\;$continuation
\cite{GujRC,Gujrati,Guj,Fedor,Corsi}.$)$ For $E=\overline{E}_{j}$, we have%

\begin{equation}
(\partial S_{j}(E)/\partial E)_{\overline{E}_{j}}=\beta,\;\;\;\overline{E}%
_{j}\in\Delta_{j}E\;\text{or }T\in\Delta_{j}T,\label{EqScond}%
\end{equation}
\ where$\ \Delta_{j}T$ is the temperature range $(T_{j},T_{j,\text{max}}),$ so
that the equilibrium energy $\overline{E}_{j}(T)$ for the basin lies in the
range $\Delta_{j}E.$ According to G2, $T_{j}$ is strictly positive for
$E_{j}=E_{\text{K}}>E_{0}$. Thus, we assert that $T_{j}$\ is not necessarily
zero \cite{GujLandscape}. The basin free energy is%

\begin{equation}
f_{j}(T)\equiv-T\ln z_{j}=\overline{E}_{j}-E_{j}-TS_{j}(\overline{E}%
_{j}),\;\;T\in\Delta_{j}T.\label{wellFrEn}%
\end{equation}
\qquad\qquad\qquad\qquad

The energy landscape is topologically very complex, with various basins very
different from each other, even if they have their minima at the same energy.
Thus, $\overline{E}_{j}(T)$ in different basins at the same temperature $T$
(provided $T\in\Delta_{j}T$ for these basins) will be usually different. There
is no requirement that they be the same. Moreover, even if the free energies
$f_{j}(T)$ of two or more basins happen to be the same at some temperature
$T$, they need not remain equal at other temperatures.

2.\qquad\textbf{Evaluating}$\ \mathit{Z}_{\lambda}\mathbf{(\mathit{T})}%
$\textbf{.} The proper form or the ISC PF's $Z_{\lambda}(T)$ or $z_{\lambda}$
in $\mathcal{I}_{\lambda}$, see (\ref{ISTotPF}), is%

\begin{equation}
Z_{\lambda}\equiv z_{_{\lambda}}e^{S_{\text{IS}}(E_{\lambda})}\equiv
\underset{j\in j(\lambda)}{\sum}z_{j}\widehat{\delta}_{T,\Delta_{j}%
T}.\label{AvBasinPF}%
\end{equation}
Due to the delta term, similar in definition to $\widehat{\delta}%
_{E,\Delta_{j}E},$ the sum in (\ref{AvBasinPF}) contains only those basins in
$\mathcal{I}_{\lambda}$ that \emph{exist} at $T$ in the sense that its range
contains $T: $ $\Delta_{j}T\ni T$. From now onward, we only consider those
basins that exist in this sense; hence, we will not explicitly exhibit the
delta term anymore. We now classify each existing basin in (\ref{AvBasinPF})
according to its free energy $f$. Let $N_{\lambda}(f)$\ denote the number of
basins of free energy $f$ that exist at a given $T$ in $\mathcal{I}_{\lambda
}.$ Since $f$ is a function of $T$, $S_{\lambda}(f)\equiv\ln N_{\lambda}(f)$
also changes with $T$. For a macroscopic system at a given fixed $T$,
$Z_{\lambda}$ is dominated by the basins in $\mathcal{I}_{\lambda}$ for which
$F_{\lambda}(f,T)\equiv f-TS_{\lambda}$ is minimum as a function of $f$
at\emph{\ fixed} $T$. The resulting entropy and the free energy at the minimum
$(f=\overline{f}_{\lambda})$ are denoted by $\overline{S}_{\lambda
}(T)=S_{\lambda}(\overline{f}_{\lambda}),$ and $\overline{F}_{\lambda
}(T)=\overline{f}_{\lambda}-T\overline{S}_{\lambda},$ respectively, where
$\overline{f}_{\lambda}$ is determined by
\begin{equation}
(\partial S_{\lambda}/\partial f)_{\overline{f}_{\lambda}}=\beta,\;\;T\geq
T_{\text{S}}^{(\lambda)}.\label{Complexity}%
\end{equation}
which looks similar to (\ref{EqScond}) but very different from
(\ref{ISTMaxCon}). Here, $T_{\text{S}}^{(\lambda)}$ is the temperature at
which $\overline{S}_{\lambda}=0,$ so that only one basin exists below it (and
above its lowest temperature $T_{j})$ in the above sense. Because of this, the
issue of $F_{\lambda}$-minimization for $\;T<T_{\text{S}}^{(\lambda)}$
\emph{does not arise} as there is only one member in $\mathcal{I}_{\lambda}$
($S_{\lambda}=0$)$,$ so that $F_{\lambda}\rightarrow$ $f_{\lambda};$ thus, the
minimization of $F_{\lambda}$ is already ensured by (\ref{EqScond}). We should
contrast (\ref{Complexity}) with (\ref{ISTMaxCon}). Stillinger puts no
restriction on the applicable temperature range in the latter; see SW2.
However, it is clear from our discussion that (\ref{ISTMaxCon}) cannot apply
below $T_{\text{SW}}$, which invalidates the second fundamental assumption
SW2. Consequently, Stillinger's argument that $T_{\text{SW}}=0$ has no
validity \cite{Stillinger1}.

3.\qquad\textbf{Evaluating}$\ \mathit{Z}\mathbf{(\mathit{T})}$. Using the
evaluated $Z_{\lambda}$ in $Z$, we find%

\begin{equation}
Z=\underset{\lambda}{\sum}e^{-\beta\lbrack E_{\lambda}+\overline{F}_{\lambda
}(T)]}.\label{Total}%
\end{equation}
We now make the following important observation. As shown above, $\overline
{f}_{\lambda}$ is independent of $E_{\lambda};$ thus, $\overline{S}_{\lambda
}(T)$ and $\overline{F}_{\lambda}(T)$ are independent of $E_{\lambda}$,
although they most certainly depend on the ISC $\mathcal{I}_{\lambda}.$ There
are various aspects of the basins such as the IS-curvature that determine
$\overline{F}_{\lambda},$ but $E_{\lambda}$ is not one of them. This has the
following very important consequence. From (\ref{AvBasinPF}), we note the
identity $f_{\lambda}\equiv$ $\overline{F}_{\lambda}+TS_{\text{IS}}%
(E_{\lambda}),$ where $f_{\lambda}(E_{\lambda},T)$ $\equiv$ $-T\ln z_{\lambda
}$ $[$ compare with $f^{(\text{SW})}(E_{\lambda},T)$ in (\ref{AvBPF})]. The
explicit dependence of $f_{\lambda}(E_{\lambda},T)$ [or $f^{(\text{SW}%
)}(E_{\lambda},T)$] on $E_{\lambda}$ must be trivial and due to $S_{\text{IS}%
}$($E_{\lambda})$ because of the \emph{independence} of $\overline{F}%
_{\lambda}$ [or $f^{(\text{SW})}(E_{\lambda},T)-TS_{\text{IS}}(E_{\lambda}%
)$]\ on $E_{\lambda}.$ Accordingly, the summation over $\lambda$ \emph{cannot}
be replaced by a summation over $E_{\lambda}.$ This disproves SW1.\ 

We now deal with the summation over $\lambda$ in (\ref{Total}) in a standard
manner$.$ Let $\mathcal{N}(\mathcal{F})$ denote the number of ISC's at a given
$T$\ with the same $\mathcal{F\equiv}E_{\lambda}+\overline{F}_{\lambda}(T).$
The PF $Z$ is dominated by the ISC's of free energy $\mathcal{F}$ for which
$\mathcal{F}-T\mathcal{S}$ is minimum over $\mathcal{F}$; here $\mathcal{S}%
(\mathcal{F})\equiv\ln\mathcal{N}(\mathcal{F}).$ The condition for this
minimum at $\overline{\mathcal{F}}$ is
\begin{equation}
(\partial\mathcal{S}/\partial\mathcal{F})_{\overline{\mathcal{F}}}%
=\beta,\;\;\;T\geq\mathcal{T},\label{ISC}%
\end{equation}
as expected. Here, $\mathcal{T}$ is the temperature at which $\overline
{\mathcal{S}}\equiv\mathcal{S}(\overline{\mathcal{F}})=0$. Hence, we finally
conclude that the final free energy is given by
\begin{equation}
F\equiv-T\ln Z\equiv\overline{\mathcal{F}}-T\overline{\mathcal{S}},\;\;\;T\geq
T_{\text{b}},
\end{equation}
where the significance of $T_{\text{b}}$ will become clear in a moment.\ It
should be obvious at this point that the dominant contribution in
(\ref{Total}) mixes ISC's with different $E_{\lambda}$ and $T_{\text{S}%
}^{(\lambda)}.$ For $T<\mathcal{T},$ the system is confined to a single ISC
corresponding a particular value $\lambda=\sigma.$ The minimization of
$\mathcal{F}-T\mathcal{S}$ is no longer an issue for $T<\mathcal{T},$ as there
is only one ISC $\sigma$ to consider and the minimization of\ $\mathcal{F}%
=\overline{F}_{\sigma}$ is already ensured by (\ref{Complexity}).

It is easy to see that the configurational entropy is
\begin{equation}
S(T)\equiv\overline{\mathcal{S}}+\overline{S}_{\sigma}+S_{\text{b}%
}.\label{Connection}%
\end{equation}
As the temperature is lowered, regardless of whether we consider CR or SCL, we
must first encounter the case $\overline{\mathcal{S}}=0$ at $\mathcal{T},$ so
that the system is confined to the single ISC $\sigma$ as discussed above. At
a lower temperature $T_{\text{S}}\equiv T_{\text{S}}^{(\sigma)}\leq
\mathcal{T}$, $\overline{S}_{\sigma}=0,$ and the system is trapped in a single
basin $j=$b. (Compare $T_{\text{S}}$ with $T_{\text{SW}}.$) The basin $j=$b
has its lowest allowed temperature $T=T_{\text{b}}<$ $T_{\text{S}}$, where the
basin entropy $S_{\text{b}}=0.$ The configurational entropy vanishes when all
three components in (\ref{Connection}) vanish. This obviously happens at
$T=T_{\text{b}}<T_{\text{S}}.$ For CR, $T_{\text{b}}=0,$ and for SCL,
$T_{\text{b}}=T_{\text{K}}>0,$ as proven recently\cite{Gujrati}. We now
restrict our discussion to SCL and consider $T_{\text{b}}\leq T<T_{\text{S}},
$ so that $\overline{S}_{\sigma}=0$ and the system is trapped in the basin
$j=$b whose IS is at $E_{\text{K}}$ and $T_{\text{b}}=T_{\text{K}}.$ In this
basin, (\ref{EqScond}) remains satisfied until $T\geq T_{\text{K}},$ but
ceases to work below $T_{\text{K}},$ if we insist that $E<$\ $E_{\text{K}}$ is
not allowed. However, if we do allow $E<$\ $E_{\text{K}},$ which requires a
mathematical continuation of SCL\ free energy function below $T_{\text{K}} $
(which is not equivalent to the continuation of the SCL\ state itself as it
does not exist below $E_{\text{K}}$)$,$ we can continue to impose
(\ref{EqScond}). This will produce a negative entropy continuation of the
SCL-entropy down to absolute zero, which is what all analytical calculations
show in which the SCL free energy\ is obtained by continuing the disordered
phase free energy below $T_{\text{M}}.$ However, demanding that (\ref{EqScond}%
) continue to operate below $T_{\text{K}},$ and at the same time demanding
that $E$ remains fixed at $E$=$E_{\text{K}}$ is \emph{not} legal. Similar
arguments apply to the consideration of $\overline{\mathcal{S}}$\ and
$T_{\text{S}}.$\ We cannot demand (\ref{Complexity}) to remain operative below
$T_{\text{S}}$; it must be replaced by (\ref{EqScond}).

It should be noted that the general landscape picture itself is not capable of
providing any information about whether $T_{\text{K}}>0$ or not. For this, we
must turn to other approaches like the one developed by Gujrati\cite{Gujrati}
or to some model landscapes. Indeed, model landscapes can be easily
constructed in which the temperature associated with the minimum of a basin
need not vanish \cite{GujLandscape}.

In summary, we have shown that when the landscape picture is carefully
developed, there is no conflict with the exact and rigorous results about
$T_{\text{K}}.$ In the process, we have also corrected some of the flaws in
the SW analysis. As the current analysis deals with the free energies and not
the energy minima, IS's play no useful role except possibly at very low
temperatures below $T_{\text{S}}$.

It is our pleasure to thank Andrea Corsi for his comments on the work.


\begin{thebibliography}{99}
\bibitem{Gujrati0}The binary mixture is described by the Ising model studied
in \cite{Gujrati} in a non-zero field.
\end{thebibliography}
\end{document}